\documentclass[3p,times,procedia]{elsarticle}
\usepackage{nupha_ecrc}

\volume{00}

\firstpage{1}

\journalname{Nuclear Physics A}

\runauth{}

\jid{nupha}

\jnltitlelogo{Nuclear Physics A}

\usepackage{amssymb}

\usepackage[figuresright]{rotating}
\usepackage[rightcaption]{sidecap}
\usepackage{amsmath}
\begin{document}

\begin{frontmatter}



\dochead{XXVIIth International Conference on Ultrarelativistic Nucleus-Nucleus Collisions\\ (Quark Matter 2018)}

\title{Measuring the Rate of Isotropization of Quark-Gluon Plasma Using Rapidity Correlations}


\author[ltu]{George Moschelli}
\author[wsu]{Sean Gavin}
\address[ltu]{Lawrence Technological University, 21000 West Ten Mile Road, Southfield, MI  48075}
\address[wsu]{Department of Physics and Astronomy, Wayne State University, Detroit, MI, 48202}

\begin{abstract}
We propose that rapidity dependent momentum correlations can be used to extract the shear relaxation time $\tau_\pi$ of the medium formed in high energy nuclear collisions. The stress-energy tensor in an equilibrium quark-gluon plasma is isotropic, but in nuclear collisions it is likely very far from this state.  The relaxation time $\tau_\pi$ characterizes the rate of isotropization and is a transport coefficient as fundamental as the shear viscosity. We show that fluctuations emerging from the initial anisotropy survive to freeze-out, in excess of thermal fluctuations, influencing rapidity correlation patterns. We show that these correlations can be used to extract $\tau_\pi$. In \cite{Gavin:2016hmv} we describe a method for calculating the rapidity dependence of two-particle momentum correlations with a second order, causal, diffusion equation that includes Langevin noise as a source of thermal fluctuations. The causality requirement introduces the relaxation time and we link the shape of the rapidity correlation pattern to its presence. Here we examine how different equations of state and temperature dependent transport coefficients in the presence of realistic hydrodynamic flow influence the estimate of $\tau_\pi$. In comparison to RHIC data, we find that the ratio $\tau_\pi/\nu \approx 5-6$ where $\nu=\eta/sT$ is the kinematic viscosity. 
\end{abstract}


\end{frontmatter}


%
%
%
\section{Introduction}
\label{sec:intro}
Values of transport coefficients indicate the characteristics of the medium produced in relativistic heavy-ion collisions. Measurements of small values for the shear viscosity are often cited as evidence that the medium is a quark-gluon plasma (QGP). Importantly, the sear viscosity alone is not sufficient to fully constrain theories of the medium. In this paper we propose a method, following \cite{Gavin:2016hmv}, for extracting another transport coefficient, the shear relaxation time $\tau_\pi$, by using the rapidity dependence of transverse momentum correlations.


It is likely that the stress-energy tensor describing QGP produced in real collisions starts far from an isotropic state.
The shear relaxation time $\tau_\pi$ characterizes the rate of isotropization of the medium due to shear viscous forces.
We demonstrate that the rapidity dependence of momentum correlations is sensitive to $\tau_\pi$ and
describe a method for calculating these correlations with a second order, causal, diffusion equation that includes Langevin noise as a source of thermal fluctuations. The causality requirement introduces the relaxation time and we link the shape of the rapidity correlation pattern to its presence. 


%
%
%
\section{Diffusion of Momentum Correlations}
\label{sec:diff}
Fluctuations in transverse momentum densities are born in the earliest moments of the collision. Effects of these fluctuations can be observable through two-particle momentum correlations if the fluctuations persist through the dynamical evolution, and final state particle momentums are representative of local momentum densities of the fluid at freeze-out. Correlations of particle momentums then mimic correlations of momentum density fluctuations in the fluid medium. These correlations can originate from several sources, including 1) initial state production mechanisms that can be long range (greater than 1-2 units of separation) in rapidity, 2) dynamical processes like diffusion that transfer momentum between fluid cells in shorter rapidity ranges, and 3)  fluctuations in the geometrical shape of the collision region. Viscous forces simultaneously destroy fluctuations while transporting momentum over space, inducing correlations between fluid cells at different positions. As viscosity drives fluid cell momentums toward an average, fluctuations and correlations decrease in magnitude but spread to larger distances. In other words, \textit{correlations} in fluctuations diffuse in the same way that concentrations of any quantity would diffuse in a fluid. Indeed texts on stochastic processes show that correlation functions diffuse in this manner \cite{gardiner2004handbook}.

We focus our interest on correlations of momentum fluctuations at two different points in the fluid
\begin{equation}\label{eq:momcorr}
r(\mathbf{x}_1, \mathbf{x}_2) = \langle \mathbf{g}(\mathbf{x}_1)\mathbf{g}(\mathbf{x}_2) \rangle - 
\langle \mathbf{g}(\mathbf{x}_1)\rangle\langle \mathbf{g}(\mathbf{x}_2)\rangle.
\end{equation}
Here $\mathbf{g}$ is the \textit{difference} in momentum density from the global average at position $\mathbf{x}$. 
More precisely, $\mathbf{g}$ is the divergence free component of the momentum density fluctuation, as will be discussed shortly. 

To calculate the evolution of correlations (\ref{eq:momcorr}), we follow \cite{Gavin:2016hmv} and use linearized, second order, M\"{u}ller-Isreal-Stewart hydrodynamics with noise and one dimensional longitudinal expansion.
The linearization builds in the assumption that fluctuations are small and fluid cells are at least near local thermal equilibrium. 
Second order hydrodynamic equations introduce a relaxation time that allows for causality to be preserved; 
changes in fluctuations at some point in the fluid cannot instantaneously influence all other points in the fluid. Importantly, this condition both introduces the transport coefficient $\tau_\pi$ into the theory and leads to the second order time derivative in the diffusion equation (\ref{eq:momdif}).
The inclusion of noise follows from the fluctuation-dissipation theorem which states that all dissipative terms are accompanied by stochastic fluctuations due to microscopic interactions. 
Finally, while shear forces cause fluid cells to transfer transverse momentum in the longitudinal direction, longitudinal expansion competes with this transfer. Therefore, gaining a realistic description of the rapidity dependence requires a description of longitudinal expansion. 

In order to investigate the evolution of momentum correlations (\ref{eq:momcorr}), we write the hydrodynamic equations for momentum density rather than velocity. 
Further, in  \cite{Gavin:2016hmv}, we show that these equations can also be written for momentum density fluctuations. 
Since we will examine transverse momentum fluctuations, we recognize that viscous forces transfer transverse momentum perpendicular to the direction of the transverse momentum of fluid cells - in the longitudinal direction.
At linear order shear and longitudinal modes are decoupled, therefore, only shear forces will contribute to the rapidity dependence of (\ref{eq:momcorr}).
Longitudinal modes may send compression sound waves in the longitudinal direction, however these modes do not change the positions of fluid cells or relative longitudinal distances between cells. Longitudinal modes also do not transfer any \textit{transverse} momentum between cells.

Given that momentum density is a vector quantity, we can separate out longitudinal and transverse modes using the Helmholtz decomposition. We indicate the divergence free component of the momentum density fluctuation as $\mathbf{g}$. Now, the correlations (\ref{eq:momcorr}) emerge from only initial state correlations and vortical forces.
By keeping only the curl free components of momentum fluctuations, longitudinal modes would modify a similar correlation function. These correlations could be used to study bulk viscosity, sound modes, and thermal and baryon conductivity. We leave this for future work.

Using only the shear components of momentum correlations we obtain a causal diffusion equation
\begin{equation}\label{eq:momdif}
\left[ \frac{\tau_\pi}{2}\frac{\partial^2}{\partial\tau^2} +
\left( 1 + \frac{\kappa\tau_\pi}{\tau} \right)\frac{\partial}{\partial\tau} -
\frac{\nu}{\tau^2}\left( 2\frac{\partial^2}{\partial(\Delta\eta)^2} + \frac{1}{2}\frac{\partial^2}{\partial\eta_a^2}\right) 
\right]\Delta r = 0,
\end{equation}
for correlations of transverse momentum density fluctuations.  We define the relative $\Delta\eta = \eta_1-\eta_2$ and average $\eta_a=(\eta_1+\eta_2)/2$ spatial rapidities for pairs of fluid cells at different longitudinal positions. Given Bjorken expansion, we equate spatial rapidity with rapidity, and $\tau$ is the proper time. 
Noise influences (\ref{eq:momdif}) in a subtle but important way. 
If viscosity has enough time, all fluid cells will approach the average value, driving $\mathbf{g}\rightarrow 0$ and $\langle \mathbf{g}(\mathbf{x}_1)\mathbf{g}(\mathbf{x}_2) \rangle\rightarrow \langle \mathbf{g}(\mathbf{x}_1)\rangle\langle \mathbf{g}(\mathbf{x}_2)\rangle$ and therefore $r(\mathbf{x}_1, \mathbf{x}_2)\rightarrow 0$. Noise causes random fluctuations - changes of $\mathbf{g}$ -  sourcing random correlations in (\ref{eq:momcorr}). Consequentially, in equilibrium, correlations (\ref{eq:momcorr}) do not vanish, but approach some value $r\rightarrow r^{eq}$. 
It turns out that the quantity $\Delta r= r - r^{eq}$ also satisfies the diffusion equation, and therefore (\ref{eq:momdif}) describes the diffusion of correlations (\ref{eq:momcorr}) \textit{in excess} of random correlations.

Three transport coefficients appear in (\ref{eq:momdif}). 
Kinematic viscosity, $\nu=\eta/sT$, scales the spatial derivatives. Here $\eta/s$ is the shear viscosity to entropy density ratio and $T$ is the temperature. The coefficient $\kappa$ scales gradients in speeds of fluid cells. Finally, the shear relaxation time $\tau_\pi=\beta\nu$ is the characteristic time that scales the time derivatives, effectively dialing the rate at which the correlation function changes. 
%
%
To focus on the effects of $\tau_\pi$, we parameterize $\eta/s$ with $T$ following \cite{Niemi:2012ry} and calculate the time dependence of temperature, entropy density, and $\kappa$ following \cite{Muronga:2001zk} (and the erratum); given Bjorken flow, causality delays heating and we use the coupled differential equations
\begin{equation}\label{eq:entropy}
\frac{ds}{d\tau} + \frac{s}{\tau} = \frac{\Phi}{T\tau}~~~~~;~~~~~
\frac{d\Phi}{d\tau}=-\frac{1}{\tau_\pi}\left(\Phi -\frac{4\eta}{3\tau} \right)-\frac{\kappa}{\tau}\Phi
~~~~~;~~~~~\kappa=\frac{1}{2}\left\{1+\frac{d\ln(\tau_\pi/\eta T)}{d\ln\tau}\right\}.
\end{equation}
%
For example, in a conformal fluid, where the only scale is $T$, $\tau_\pi\sim1/T$, $\eta\sim s\sim T^3$, and $\kappa=4/3$.
\section{Rapidity Correlations}
\label{sec:corr}
\begin{figure}[ht]
	\centering
	\includegraphics[width=0.7\textwidth]{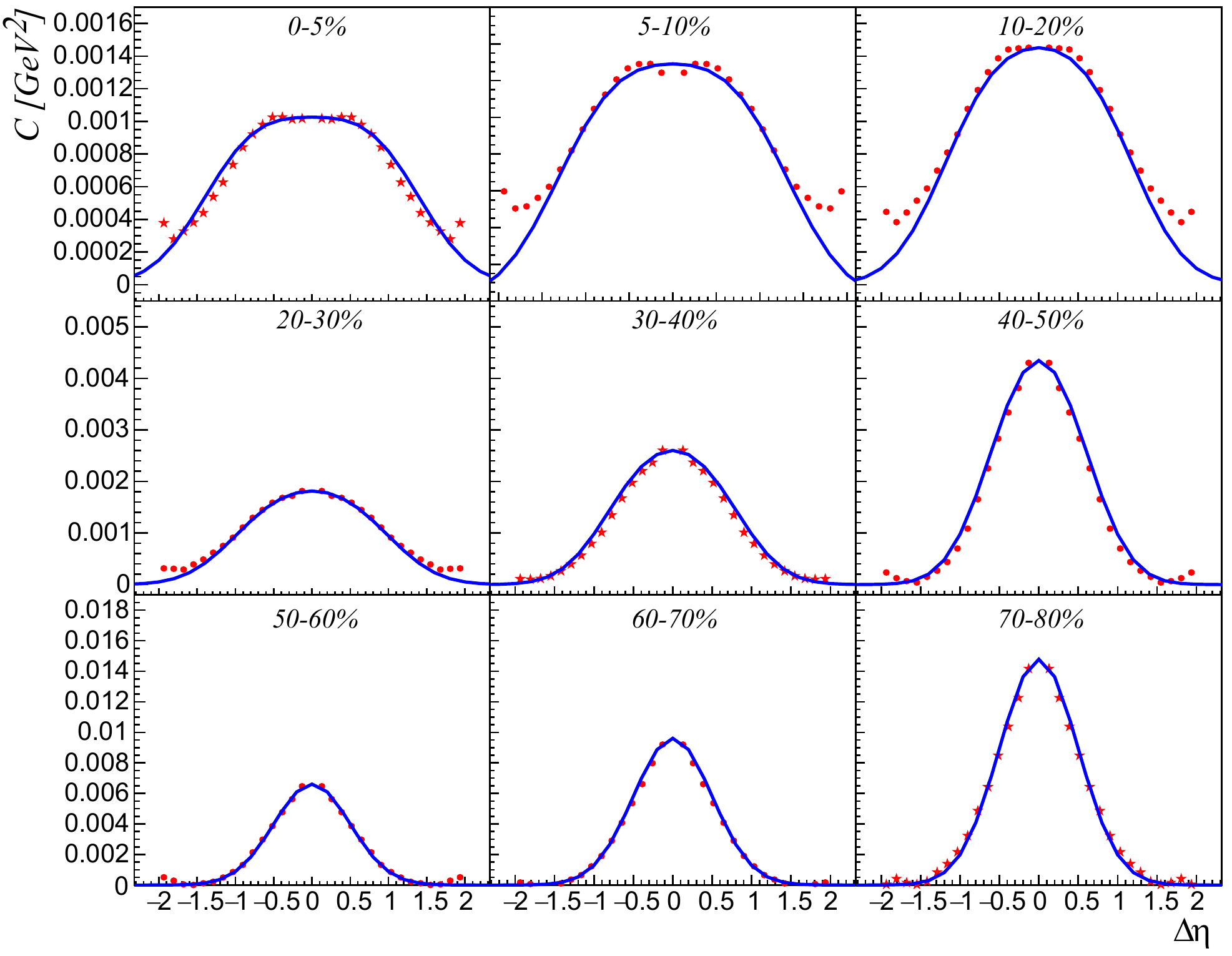}
	\caption{\label{fig:uyr} Relative rapidity distribution of transverse momentum correlations for Au+Au collisions at $\sqrt{s}=200~GeV$. $\beta=5.5$, $\tau_0 = 1.05~fm$, $\tau_{f,b=0} = 1.05~fm$, $T_{FO}=150~MeV$. 
	Data is from the STAR experiment \cite{Agakishiev:2011fs} }
\end{figure}

Correlations (\ref{eq:momcorr}) are observable by measuring two-particle momentum correlations,
\begin{equation}\label{eq:C}
{\cal C} = \langle N\rangle^{-2}\left\langle \sum\limits_i\sum\limits_{j\neq i}p_{t,i}p_{t,j}\right\rangle - 
\langle p_t\rangle^2
= \langle N\rangle^{-2} \iint \Delta r(\mathbf{x}_1, \mathbf{x}_2) d\mathbf{x}_1 d\mathbf{x}_2
.
\end{equation}
where $p_{t}$ is the transverse momentum of particle $i$ or $j$ and $\langle ...\rangle$ is an event average.
The STAR experiment has measured (\ref{eq:C}) differentially as ${\cal C}(\Delta\eta,\Delta\phi)$ where $\Delta\eta$ is the relative pseudorapidity and $\Delta\phi$ is the relative azimuthal angle between particle pairs \cite{Agakishiev:2011fs}. A differential measurement in $\Delta\phi$ can be used to study the contributions from geometrical fluctuations and we leave this for future work. To eliminate contributions to (\ref{eq:momcorr}) from geometrical fluctuations we integrate ${\cal C}$ over $\Delta\phi$. Since geometrical fluctuations are characterized by a Fourier cosine series, the integral of any $\cos(n\Delta\phi)$ 
on the interval $0<\Delta\phi<2\pi$ 
will vanish. 
In Fig. \ref{fig:uyr}, STAR integrated ${\cal C}(\Delta\eta,\Delta\phi)$ in the region $|\Delta\phi|<1$.

Figure \ref{fig:uyr} shows the comparison of momentum correlations (\ref{eq:C}) calculated from (\ref{eq:momdif}) using a parameterization of $\eta/s$ from \cite{Niemi:2012ry} with the time dependence of parameters found using (\ref{eq:entropy}). We found best agreement using a formation time of $\tau_{0} = 1.05~fm$, a freeze-out temperature of $T_{FO}=150~MeV$ and  $\beta=\tau_\pi/\nu=5.5$. The value of $\beta=5.5$ and the non-Gaussian shape of the distribution are our primary results.

The departure from a Gaussian form of the rapidity profile of ${\cal C}(\Delta\eta)$ is a consequence of enforcing causality. Examining (\ref{eq:momdif}), one can notice a resemblance to the Telegrapher's equation which has wave and diffusion components. 
The distribution of correlations $\Delta r$ in relative rapidity $\Delta\eta$ is initially a single peak that separates into a pair of oppositely moving wave pulses with a diffusion hump filling the region between them. The pulses move with a wave speed $\beta^{-1/2}$ in Cartesian space, but longitudinal expansion slows the propagation in rapidity by a factor of $\tau^{-2}$. Freeze-out happens rapidly such that the wave pulses separate only enough to produce a flattening of the profile near $\Delta\eta=0$. The data suggests a dip at $\Delta\eta=0$ in central collisions which we interpret as the wave pulses separating. It may be possible to resolve the dip with better knowledge of the initial distribution of correlations.
We chose the initial width of $\Delta r$ to match distribution in the most peripheral collisions, but it is likely that the initial distribution is narrower.

%
%
%
\section{Conclusion}
\label{sec:concusion}
Using the rapidity dependence of two-particle momentum correlations (\ref{eq:C}), we are able to extract a value of the shear relaxation time, $\tau_\pi$, a transport coefficient that characterizes the rate of isotropization of the medium produced in relativistic heavy-ion collisions. 
We follow \cite{Gavin:2016hmv}, and outline a method for calculating the rapidity dependence of (\ref{eq:C}) with a second order, causal, diffusion equation that includes Langevin noise as a source of thermal fluctuations. We find good agreement with data using $\tau_\pi/\nu = 5.5$, where $\nu=\eta/sT$ is the kinematic viscosity. Significantly, we obtain comparable results using the range $5 <\tau_\pi/\nu <6$. This indicates that the data can indeed be used to put realistic limits on the value of $\tau_\pi$.

One significant benefit of this work is the development of evolution equations for \textit{correlations} rather than point-by-point densities on a grid. 
In contrast to hydrodynamic or transport simulations that have computationally expensive numerics and lose two-particle spatial correlation information when performing freeze-out, we map the evolution of correlations throughout the whole lifetime of the collision with comparably small numerical resources.
Finally, we comment that more information can be obtained about two-particle correlation functions by studying pre-equilibrium fluctuations \cite{Gavin:2016nir}.
\vspace{-0.3cm}
\section*{Acknowledgments}
\vspace{-0.2cm}
The authors would like to give special thanks to Tyler Kostun and Christopher Zin. 
This work was supported in part by the U.S. NSF grant PHY-1207687.
\vspace{-0.3cm}


\bibliographystyle{elsarticle-num}
\bibliography{references}

\begin{thebibliography}{1}
\expandafter\ifx\csname url\endcsname\relax
  \def\url#1{\texttt{#1}}\fi
\expandafter\ifx\csname urlprefix\endcsname\relax\def\urlprefix{URL }\fi
\expandafter\ifx\csname href\endcsname\relax
  \def\href#1#2{#2} \def\path#1{#1}\fi

\bibitem{Gavin:2016hmv}
S.~Gavin, G.~Moschelli, C.~Zin, {Rapidity Correlation Structure in Nuclear
  Collisions}, Phys. Rev. C94~(2) (2016) 024921.
\newblock \href {http://arxiv.org/abs/1606.02692} {\path{arXiv:1606.02692}},
  \href {http://dx.doi.org/10.1103/PhysRevC.94.024921}
  {\path{doi:10.1103/PhysRevC.94.024921}}.

\bibitem{gardiner2004handbook}
C.~Gardiner, Handbook of Stochastic Methods for Physics, Chemistry, and the
  Natural Sciences, Springer, 2004.

\bibitem{Niemi:2012ry}
H.~Niemi, G.~S. Denicol, P.~Huovinen, E.~Molnar, D.~H. Rischke, {Influence of a
  temperature-dependent shear viscosity on the azimuthal asymmetries of
  transverse momentum spectra in ultrarelativistic heavy-ion collisions}, Phys.
  Rev. C86 (2012) 014909.
\newblock \href {http://dx.doi.org/10.1103/PhysRevC.86.014909}
  {\path{doi:10.1103/PhysRevC.86.014909}}.

\bibitem{Muronga:2001zk}
A.~Muronga, {Second order dissipative fluid dynamics for ultrarelativistic
  nuclear collisions}, Phys.Rev.Lett. 88 (2002) 062302.
\newblock \href {http://dx.doi.org/10.1103/PhysRevLett.88.062302}
  {\path{doi:10.1103/PhysRevLett.88.062302}}.

\bibitem{Agakishiev:2011fs}
H.~Agakishiev, et~al., {Evolution of the differential transverse momentum
  correlation function with centrality in Au$+$Au collisions at $\sqrt{s_{NN}}
  =$ 200 GeV}, Phys. Lett. B704 (2011) 467--473.
\newblock \href {http://arxiv.org/abs/1106.4334} {\path{arXiv:1106.4334}},
  \href {http://dx.doi.org/10.1016/j.physletb.2011.09.075}
  {\path{doi:10.1016/j.physletb.2011.09.075}}.

\bibitem{Gavin:2016nir}
S.~Gavin, G.~Moschelli, C.~Zin, {Boltzmann-Langevin Approach to Pre-equilibrium
  Correlations in Nuclear Collisions}\href {http://arxiv.org/abs/1612.07856}
  {\path{arXiv:1612.07856}}.

\end{thebibliography}

\end{document}